\documentstyle[preprint,eqsecnum,aps,psfig]{revtex}
\tightenlines

\begin{document}
\draft
\title{Orientation of Vortices in a Superconducting Thin-Film: Quantitative 
Comparison of Spin-Polarized
Neutron Reflectivity and Magnetization}
\author{S.-W. Han$^\dagger$$^*$, J. Farmer$^\ddagger$, H. Kaiser$^\ddagger$, 
P. F. Miceli$^\dagger$$^{**}$}
\address{$^\dagger$Department of Physics and Astronomy,
University of Missouri-Columbia, Columbia, Missouri 65211}
\address{$^\ddagger$Missouri University Research Reactor, Columbia,
Missouri 65211}
\author{I. V. Roshchin, L. H. Greene}
\address{Department of Physics, University of Illinois at
Champaign-Urbana, Urbana, Illinois 61801}
\date{\today}
\maketitle

\begin{abstract}

We present a quantitative comparison of the magnetization measured by 
spin-polarized neutron reflectivity (SPNR) and DC magnetometry 
on a 1370 \AA\ -thick Nb superconducting film. 
As a function of magnetic field applied in the film plane, SPNR exhibits
reversible behavior whereas the DC magnetization shows substantial hysteresis. 
The difference between these measurements is attributed to a 
rotation of vortex magnetic field out of the film plane as the applied field is reduced. 
Since SPNR measures only the magnetization parallel to the film 
plane whereas DC magnetization is strongly influenced by the 
perpendicular component of magnetization when there is a slight sample 
tilt, combining the two techniques allows one to distinguish two components 
of magnetization in a thin film. 

\end{abstract}
\vskip 10in
\section{Introduction}

The current transport properties of type II superconductors depend 
critically on the mobility of vortices and, therefore, the subject of vortex 
pinning has attracted considerable attention \cite{blatter}.
Since the addition or removal of a vortex requires an encounter with the 
superconductor's surface, the interaction of vortices with surfaces and 
interfaces play an important role in the current transport properties \cite{fuchs,zeldov}. 
For thin-film superconductors, the surface has a substantial effect 
on vortex behavior. 
In this case, a large anisotropy between vortices 
parallel or perpendicular to the film plane might be anticipated. 
For example, recent studies of artificially layered superconducting thin films 
suggest that vortices parallel to the surface can order spatially due to the 
interaction with the surface \cite{guimpel}. 

In studies of thin-film superconductors it is useful to apply 
the magnetic field parallel to the surface. 
This geometry can, however, be problematic for experiments. 
A magnetization measurement presents 
special difficulties due to the large demagnetization factor 
perpendicular to the surface combined with a small sample tilt angle 
that is unavoidable.
On the other hand, local probe techniques, such as the Bitter method, 
scanning probe techniques, and Lorentz microscopy, are constrained to 
measure the magnetic flux as it emerges through the surface (i.e. 
vortices are perpendicular to the surface). 
Alternatively, small angle neutron diffraction methods require large 
volume samples so that thin films cannot be studied. 
Since the selection rules for the specular reflection of 
spin-polarized neutrons from a thin film guarantees that only the 
magnetization parallel to the surface is measured,
spin-polarized neutron reflectivity (SPNR) should be useful for 
measuring the magnetization of thin-film superconductors.

The application of SPNR to 
thin-film superconductors was demonstrated by G. Felcher {\it et 
al.} \cite{felcher} where the London penetration length, $\lambda_L$, was 
measured for Nb. 
Further refinement of this technique for Nb by H. Zhang {\it et al.} 
\cite{zhang} included the effect of the electron mean-free path on 
$\lambda_L$.
The same technique has also been used to determine $\lambda_L$ for 
high-temperature superconducting oxides \cite{felcher2}.
All of these studies were performed at low field, below H$_{c1}$. 
Recently, Han {\it et al.} \cite{han} demonstrated that SPNR is also 
useful at higher fields, above H$_{c1}$, where it can yield information on 
vortices. 
In particular, the average vortex density was extracted for the field 
parallel to a thin film of YBa$_2$Cu$_3$O$_{\rm 7-x}$ (YBCO) and a large 
hysteresis loop was observed.

In this paper, we compare SPNR and DC magnetization measurements on a Nb 
thin film for the field applied parallel to the surface.  The two 
measurements exhibit different field dependences: hysteresis is observed 
for the DC magnetization whereas the SPNR is reversible.   
The difference is attributed to the magnetic field of vortices that is
generated perpendicular
to the film plane as the applied field is decreased.  
By quantitatively comparing the SPNR and magnetization measurements
we have deduced the rotation of the magnetic field as a function of 
the applied field.

\section{Experimental Details}

A Nb/Al thin-film bilayer was grown by sputter-deposition on a 1 cm$^2$ 
(1$\bar 1$02) single-crystal Al$_2$O$_3$ substrate \cite{greene1}. 
A 50-hour pumpdown in a UHV-compatible stainless-steel vacuum chamber was preformed with simultaneous substrate baking at 520$^o$C.  
Before deposition a cold trap was filled with liquid nitrogen, providing a base pressure of less than 1x10$^{-8}$ Torr.  
Sputter deposition was performed under a 7x10$^{-3}$ Torr atmosphere of Ar.  
With the substrate temperature held at 470$^o$C,  a 1370 \AA\ -thick layer of Nb was deposited at a rate of 8 \AA /sec.  
Without breaking vacuum, the substrate temperature was then reduced to 120$^o$C
and a 760 \AA\ -thick layer of Al was deposited {\it in-situ} on the Nb surface. 
The superconducting transition temperature was determined to be 
T$_c$ = 9.25 $\pm$ 0.25 K 
from the magnetization measured at 50 Oe, 
which is consistent with T$_c$ =  
9.11 K (transition width = 0.06 K) found by 4-probe resistivity 
measured on an adjacently grown sample.  
The residual resistivity ratio was measured to be
$\rho_{\rm 300 K}/\rho_{\rm 10 K} = 25.4$.

The SPNR measurements were performed using  
the GANS reflectometer \cite{kaiser} at the Missouri University Research 
Reactor (MURR) with a neutron wavelength of 2.35 \AA. 
A supermirror was employed to select the spin state of the neutron beam to 
be perpendicular to the scattering plane and the spin state could be 
switched using a flipper coil located downstream from the mirror. 
A polarization analyzer was not used.
The beam width at the sample position, $\simeq$0.23 mm, and the angular
divergence of the beam, 0.018$^o$, were measured in the scattering 
plane.
During the data analysis, the instrumental resolution as well as 
the incident beam polarization efficiency, 93\%,  
were taken into account.   
The specular reflectivity was consecutively measured for spin-up and 
spin-down neutrons near the critical angle for total external reflection.
The sample was mounted in a cryostat having
a superconducting split-coil magnet where the
field was applied perpendicular to the scattering plane and parallel to the 
sample surface. 
The sample was always cooled to 2 $\pm$ 0.2 K in zero magnetic field 
before a field was applied.  

A Quantum Design 
superconducting quantum interference device (SQUID) magnetometer was used
for the magnetization measurements which were performed on the 
same sample used for the neutron experiments.  This sample was cut to 0.36 
$\times$ 0.26 cm$^2$ and placed in the
magnetometer cryostat such that the sample surface was near-parallel to 
the applied field.
The tilt angle between the sample surface and the applied field was set when it 
was mounted in the magnetometer.  
The angular reproducibility was $\pm$ 0.25$^o$ and the zero angle was determined
from a fit to the angular dependence of the DC susceptibility measured at
low field.  
The reported magnetization curves were obtained for the sample cooled in zero
magnetic field. 

\section{Results}

\begin{figure}
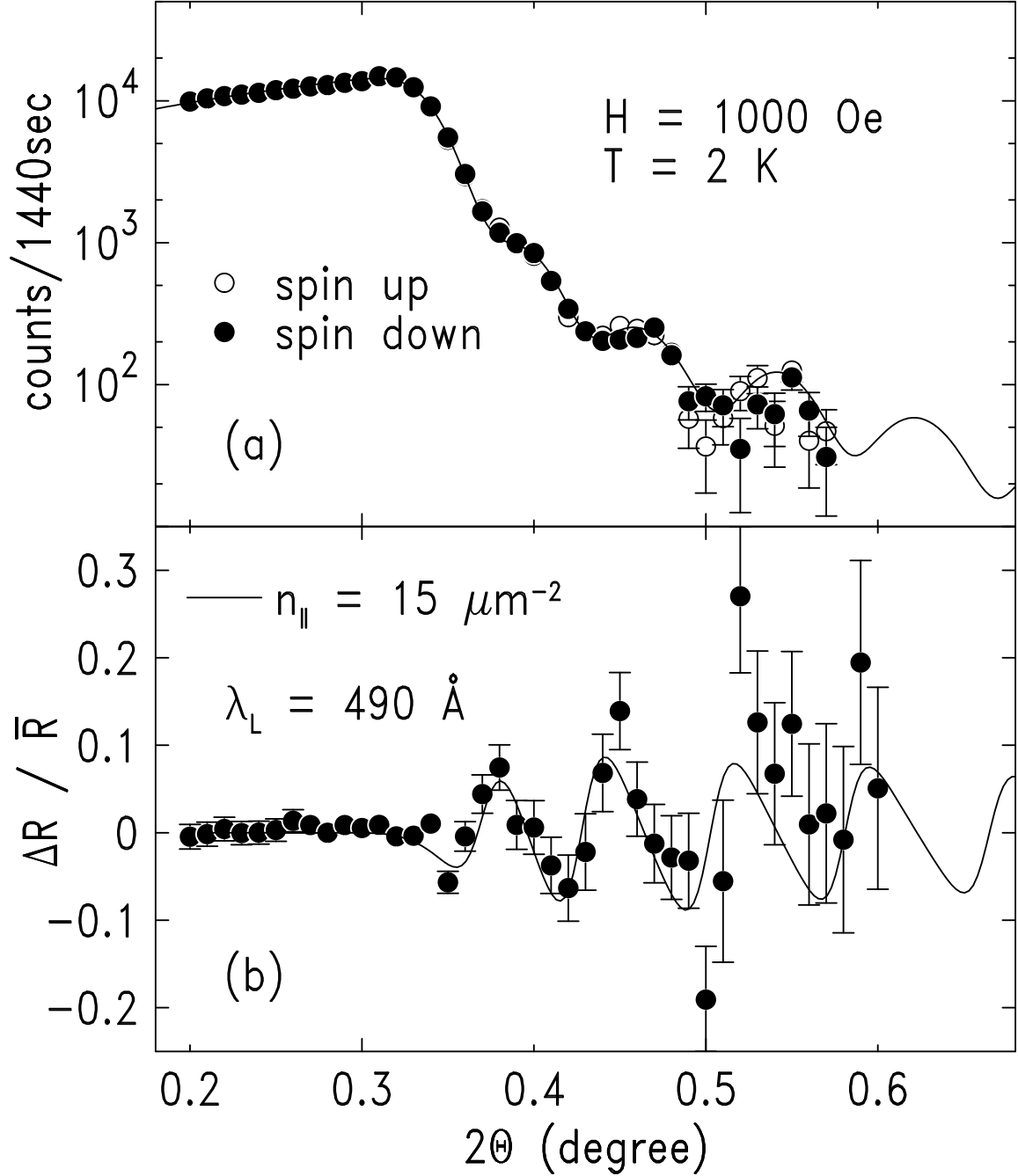

\caption{(a) The number of counts are measured as a function of scattering
angle for spin-up and spin-down 
neutrons reflected from a Al/Nb/Al$_2$O$_3$ film at 2 K and 1000 Oe. 
(b) The difference between the spin-up and down 
reflectivity divided by the average reflectivity for the two spin states,
$\Delta R/\overline{R}$, is obtained from the data in (a).
The solid curves are calculated using the model of 
ref. [8]
with a magnetic screening length of 490 \AA\ and uniform vortex density of
15 $\mu m^{-2}$.}
\label{fig1}
\end{figure}

The number of specularly reflected spin-up and 
spin-down neutrons measured as a function of the scattering angle, 
2$\theta$, for the sample held at 2 K and 1000 Oe is shown in Fig.\ \ref{fig1} (a). 
The interference fringes arise from the finite thickness of the film. 
A least squares fit using a reflectivity model \cite{russell} 
yields the Nb film thickness (1370 $\pm$ 10 \AA) as well as the Al/Nb (30 
$\pm$ 10 \AA) and Nb/Al$_2$O$_3$ (2 $\pm$ 1 \AA) interface roughnesses.  
Since the neutron scattering cross section of Al is small relative to Nb and the neutron
absorption in Al is negligible, the Al layer had little effect on the measurement. 
X-ray specular reflectivity, measured on the same sample, gave an Al thickness of 
760 $\pm$ 30 \AA, a roughness of 70 $\pm$ 16 \AA\ at the air/Al interface, and the other 
quantities were consistent with the neutron measurement.
The magnetic contribution to the neutron reflectivity can be observed by plotting the 
difference between the spin-up and the spin-down 
reflectivities, $\Delta R$, divided by the average of the spin-up and down
reflectivity, $\overline{R}$, as shown in Fig.\ \ref{fig1} (b).

\begin{figure}
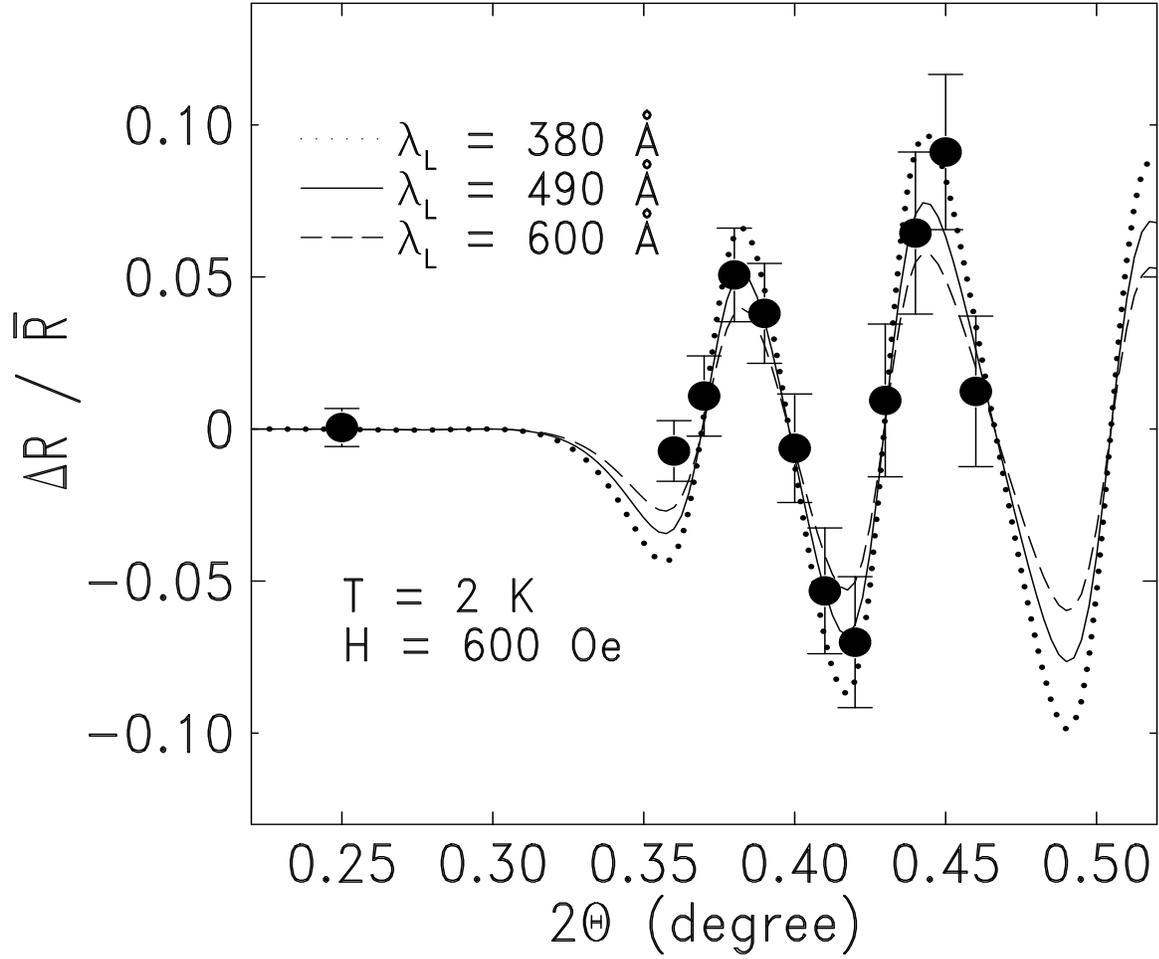

\caption{The London penetration length was determined to be
$\lambda_L$ = 490 $\pm$ 110 \AA\ by measuring
$\Delta R/\overline{R}$ as a function of angle at 600 Oe and 2 K where the 
vortex density is negligible.
The curves are calculated for different 
screening lengths without including vortices.}
\label{fig2}
\end{figure}

The London penetration length, $\lambda_L$, for this sample was obtained from a
SPNR measurement at small applied field (600 Oe), as shown in Fig.\ \ref{fig2}.
Since the contribution of 
vortices to the magnetization is negligible at this field, 
the data were fit \cite{han} assuming no vortices.  This yields
$\lambda_L$ = 490 $\pm$ 110 \AA, in agreement with 
previous work \cite{felcher,zhang}.

\begin{figure}
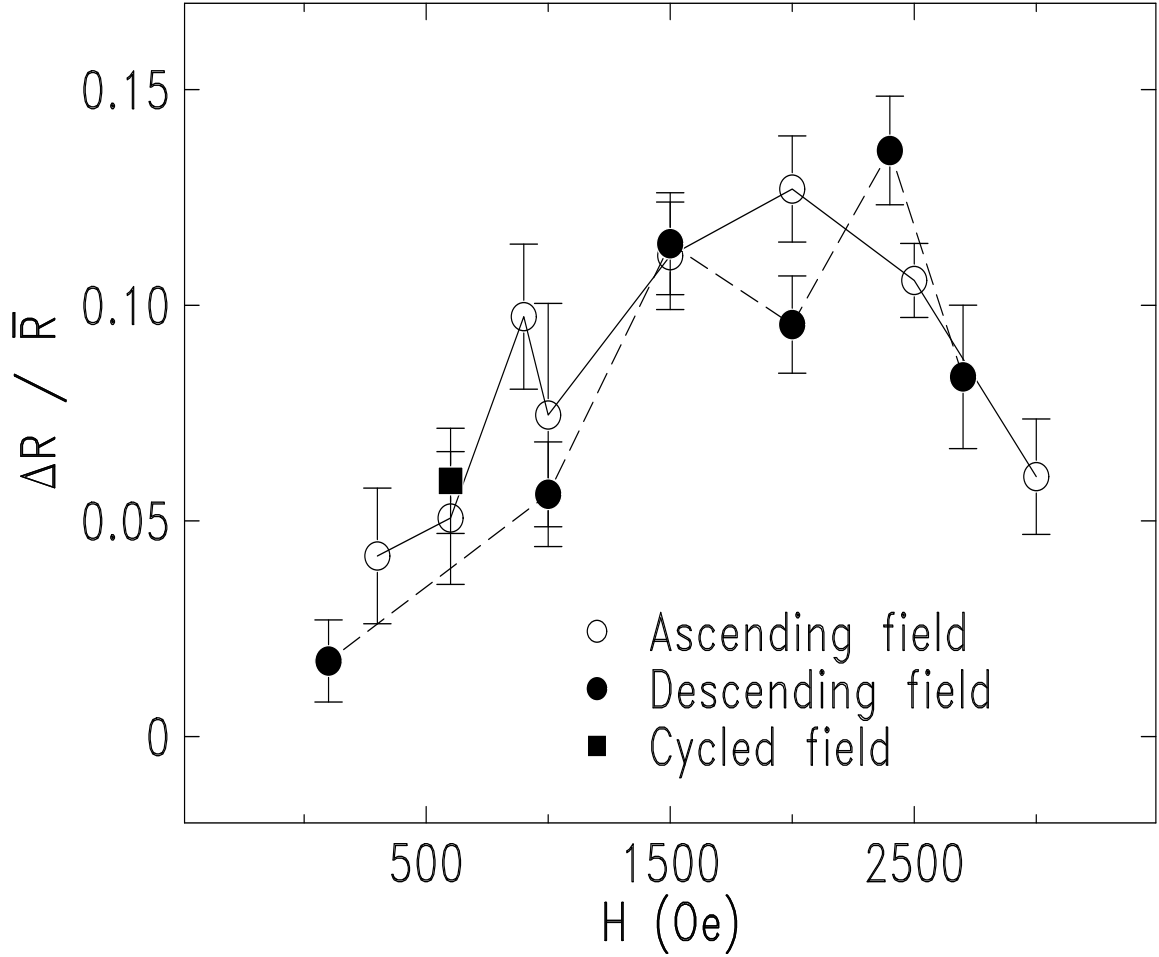

\caption{The field dependence of $\Delta R/\overline{R}$ was
measured at a single angle, 2$\theta$ = 0.38$^o$, 
for ascending and descending field at 2 K.
The lines are a "guide to the eye".
London screening dominates at low field but above $\sim$1000 Oe the introduction 
of vortices leads to saturation and the eventual reduction 
of $\Delta R/\overline{R}$.
No hysteresis is observed upon reducing the field.
The data point given by the solid square was obtained after cycling 
the applied field from 0 to 3000 Oe to -3000 Oe and finally to 600 Oe, 
which was the field during the mearurement.}
\label{fig3}
\end{figure}

In order to observe the effect of vortices, the reflectivity was 
measured as a function of applied field at a single position of 
2$\theta$ = 0.38$^o$, where $\Delta R$/$\overline{R}$ exhibits the first maximum.
Figure\ \ref{fig3} shows the
results for ascending and descending fields. 
With ascending field (open circles), it is seen that $\Delta R/\overline{R}$ 
increases steadily until $\sim$1500 Oe where it 
begins to decrease as large numbers of vortices enter the sample. 
It can also be seen that the data for descending field follow the 
ascending field data and there is no remanence at low field.
This reversible behavior is quite different from the previous 
SPNR study of a YBCO film where 
$\Delta R/\overline{R}$ exhibited a large hysteresis loop \cite{han}. 

\begin{figure}
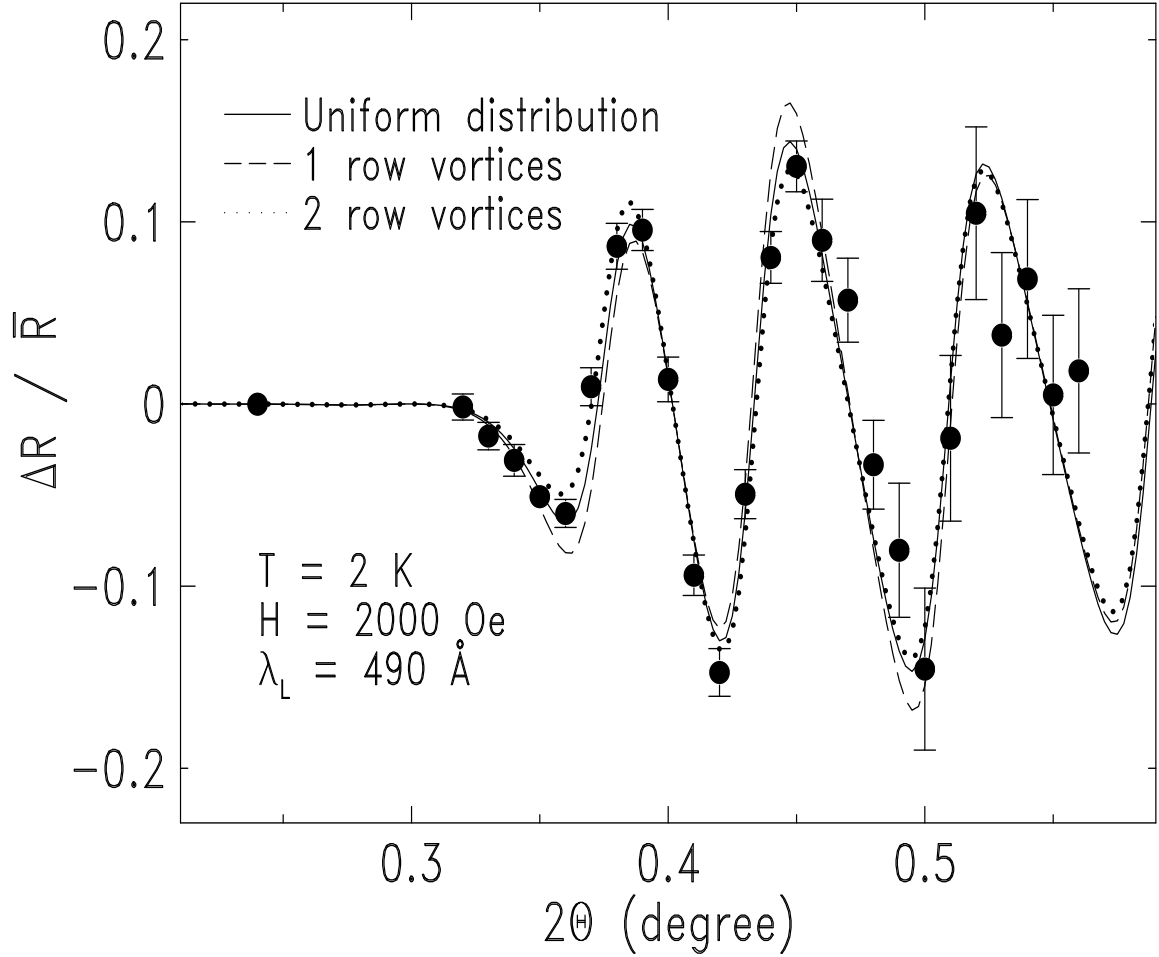

\caption{$\Delta R/\overline{R}$ is measured as a function of angle
at 2000 Oe and 2 K after reducing the field from 3000 Oe.
It is shown that the analysis is independent of the 
particular choice of vortex distribution. 
The curves are best fits for different spatial distributions of vortices:
uniform distribution with n$_\parallel$ = 40 $\mu m^{-2}$, 
vortices in a single row at the center of the film
with n$_\parallel$ = 28 $\mu m^{-2}$, and vortices in a 
double row about the center of the film with n$_\parallel$ = 33 $\mu m^{-2}$.} 
\label{fig4}
\end{figure}

The data were analyzed to extract the vortex density using a model 
presented previously \cite{han}.
Although this requires the spatial distribution of the 
vortices as input information, the analysis of the present experiments does not 
depend on the choice of distribution, as shown in Fig.\ \ref{fig4} for
data taken at 2000 Oe after the sample had been subjected to a field 
of 3000 Oe.  The three fitted curves 
correspond to different spatial distributions:  a uniform 
distribution (solid curve) with a vortex density of 40 $\mu m^{-2}$, 
vortices arranged in a single row in the middle of the film (dashed 
curve) with a density of 28 $\mu m^{-2}$, and vortices arranged in a 
double row (dotted curve) with a density of 33 $\mu m^{-2}$.  The given 
densities were chosen to yield a best fit for each spatial distribution. 
There is only a small difference between the curves because the 
interference between the reflections from the front and back interfaces 
of the superconducting film dominates
over the contribution from the vortex spatial distribution. 
This effect depends on the relative nuclear scattering lengths of the layers 
and was not observed previously for YBCO/SrTiO$_3$.

\begin{figure}
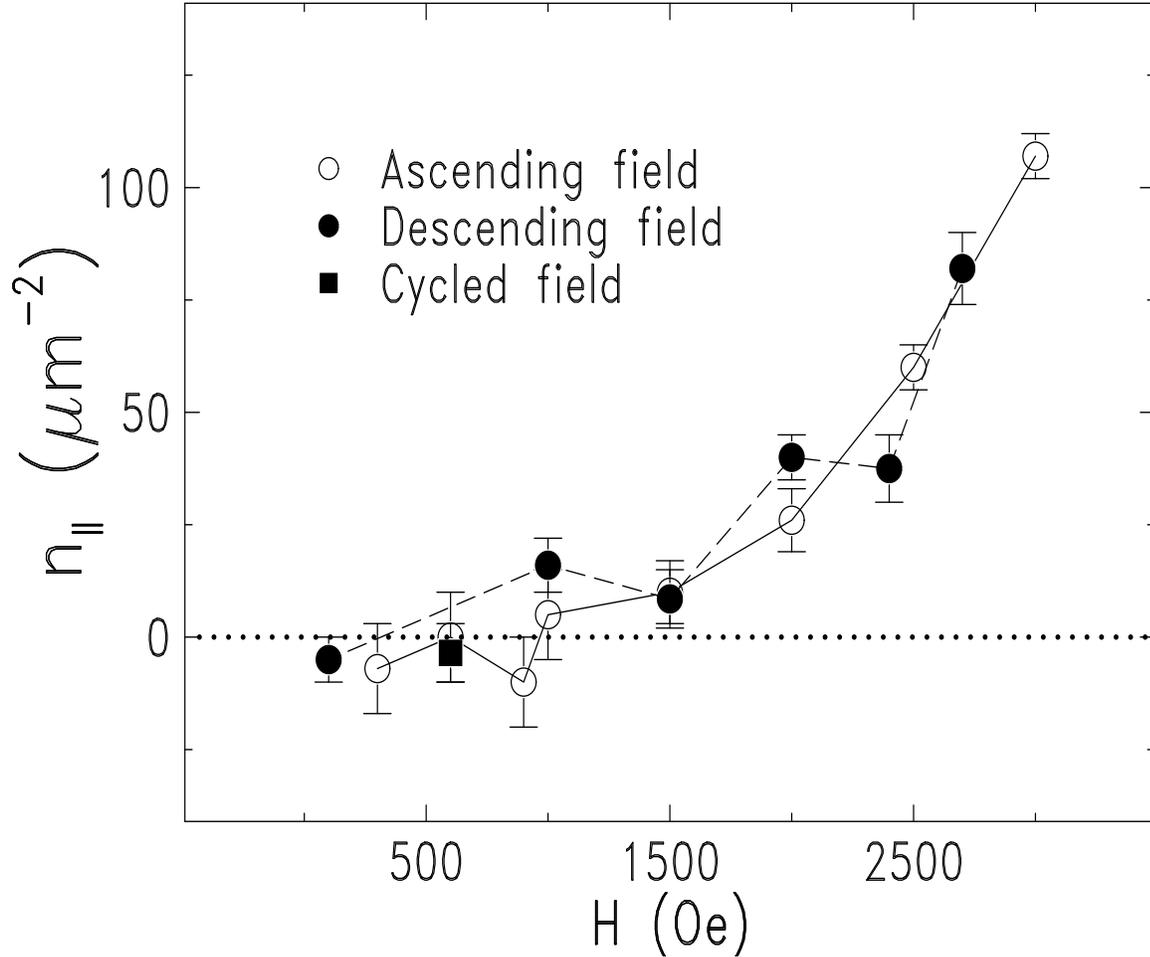

\caption{The vortex densty n$_\parallel$ as a function of applied field
is determined at 2 K from the SPNR data in 
Fig.\ \ref{fig3} assuming a uniform spatial distribution of vortices, 
$\lambda_L$ = 490 \AA, and the model of 
ref. [8].
These results show that n$_\parallel$ is independent of whether 
the applied field is ascending or descending.
The data point given by the solid square was obtained after cycling 
the applied field from 0 to 3000 Oe to -3000 Oe to 600 Oe.}
\label{fig5}
\end{figure}

Using the magnetic screening length $\lambda_L$ = 490 \AA\ and a 
uniform distribution of vortices in the model of ref. \cite{han}, the 
average vortex density parallel to the surface, n$_\parallel$, was 
determined from the data of Fig.\ \ref{fig3}.
The result, shown in Fig.\ \ref{fig5}, gives H$_{c1}$ $\sim$1000 Oe 
and the vortex density is found to increase steadily at higher field. 
Also, n$_\parallel$ exhibits reversible behavior. 
The full angle-dependent curves for $\Delta R/\overline{R}$ are calculated and compared with
the data in Figs.\ \ref{fig1} (b),\ \ref{fig2}, and\ \ref{fig4} --  
the good agreement demonstrates that it is sufficient to measure
$\Delta R/\overline{R}$ at a single angle in order to obtain n$_\parallel$. 

\begin{figure}
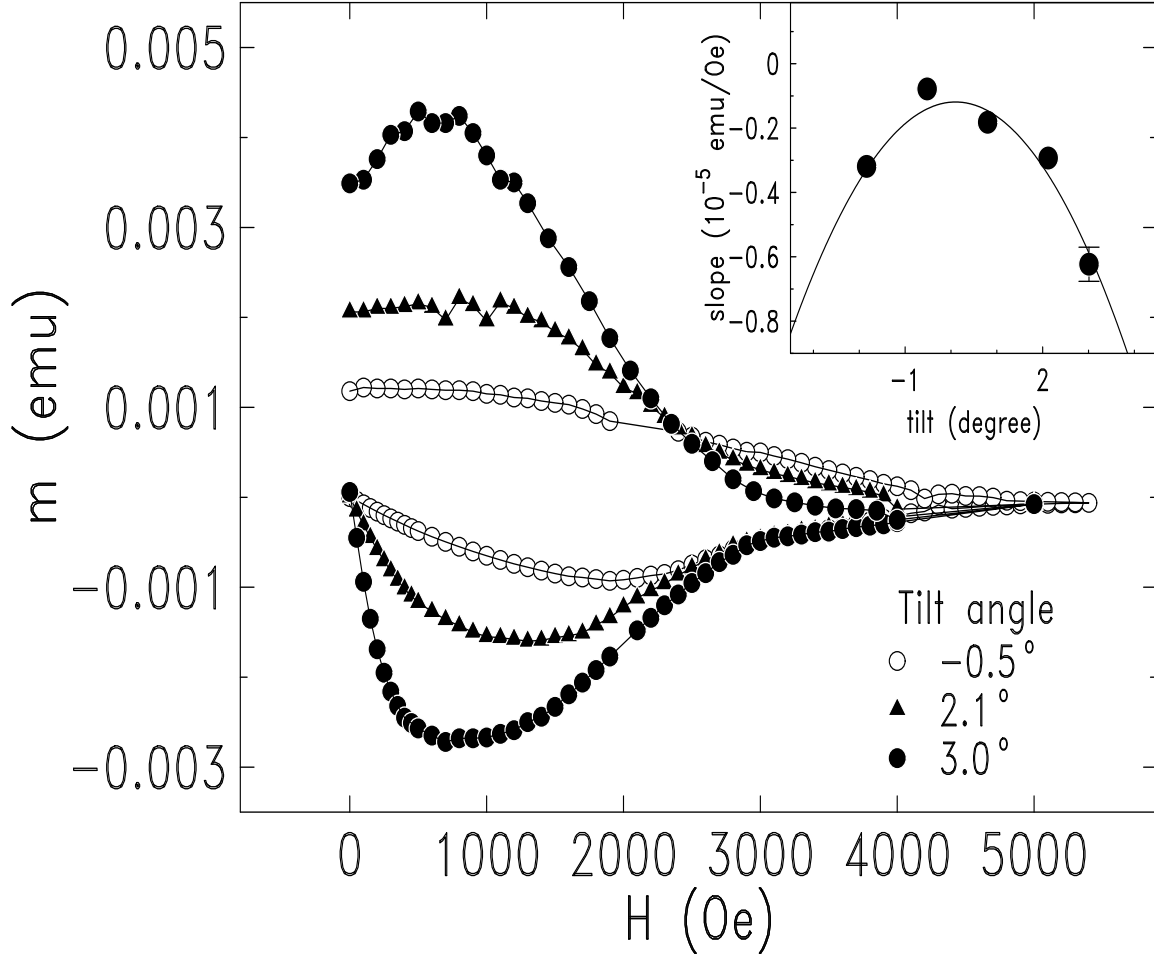

\caption{The magnetic moment determined by the SQUID magnetometer and measured 
as a function of applied field at 2 K exhibits hysteresis.  
The arrows indicate the order in which the data were collected, starting from an 
initially zero-field-cooled sample. 
These results depend on the angle of tilt between the applied field direction and the film plane.  
The inset shows the initial slope 
(linear region without vortices at small field) 
of the magnetic moment as a function of the tilt angle. 
Demagnetization effects quantitatively explain the low-field angle 
dependence and the solid curve is a best fit to Eq.\ \ref{eq5}.}
\label{fig6}
\end{figure}

We now discuss magnetization measurements that provide
additional information on the vortex behavior.  
The SQUID magnetization measured as a function of 
applied field for different tilt angles is shown in Figure\ \ref{fig6}.  
For each angle, the sample was first zero-field cooled to 2 K.
The magnetization was subsequently measured from zero to the maximum 
field and then to zero field. 
Above H$_{c1}$ the magnetization curves do not show the maxima that have 
been observed in some artificially layered superconductors 
\cite{guimpel}, suggesting there are no transitions of the 
vortex distribution in the film. 
These data, as well as other measurements performed at higher field (not shown), permit
an estimate of the upper critical field H$_{c2} \sim$5300 $\pm$ 300 Oe. 
From this, the coherence length can be calculated: $\xi$ = $\sqrt{\Phi_o / (2\pi 
H_{c2})}$ = 250 $\pm$ 8 \AA.  
This is somewhat smaller than the 390 \AA\ obtained for Nb by Pronin {\it
et al.} who also reported the London penetration depth to be 350 \AA \cite{pronin}.
Since the samples were grown in the same sputtering system 
using the same procedure, the difference might arise from 
the fact that the two techniques determine the coherence length in somewhat 
different ways.
Using the magnetic screening length (490 \AA) and coherence 
length (250 \AA), we estimate the lower critical field 
in the thin film limit \cite{abrikosov}, 
H$_{c1}$ = $\Phi_o K_0(\xi / \lambda_L) / (4 \pi \lambda_L^2) \times (1 + 2e^{-t/2\lambda_L})$,
to be $\sim$1000 Oe, which
agrees well with both the SPNR (Fig.\ \ref{fig5}) and the magnetization 
measurements (from the smallest tilt angle of -0.5$^o$ in Fig.\ \ref{fig6}).

The most striking features of the SQUID data are the
large hysteresis, in sharp contrast to the SPNR measurement, as 
well as the strong dependence on tilt angle. 
Both features are interrelated and we first discuss the effect of angle.

If a magnetic field, H$_\perp$, is applied perpendicular to the surface 
of a superconductor then the magnetization perpendicular to the surface 
M$_\perp$ is strongly affected by the geometry according to 
\cite{abrikosov,malozemoff}:
\begin{equation}
 4 \pi M_\perp = - { H_\perp - n_\perp \Phi_o \over (1 - N)},
\label{eq1}
\end{equation}
where N is the demagnetization factor, n$_\perp$ is the 
component of vortex density 
oriented perpendicular to the surface, and $\Phi_o$ is the flux
quantum, hc/2e = 20.679\  G\ $\mu$m$^2$. 
Note that n$_\perp \Phi_o$ is just the component of magnetic field 
perpendicular to the film plane.  
For a field applied parallel to the surface of a 
superconductor, assuming a uniform distribution of vortices, the 
magnetization parallel to the surface M$_\parallel$ is obtained by 
direct integration of the equations in ref. \cite{han}:
\[4 \pi M_\parallel = {1 \over t} \int_{-t/2}^{t/2} \left [H_\parallel
\left\{{cosh(z / \lambda_L) \over
cosh(t / 2\lambda_L)} - 1 \right\} \right. \]
\[ \left.+ { n_\parallel \Phi_o \over 2 \lambda_L} \int_{-t/2}^{t/2}  
\left\{ e^{-|z - z'| / \lambda_L}
- e^{(z - t/2) / \lambda_L}{sinh([z' + t/2] / \lambda_L) \over sinh(t
/ \lambda_L)}
+ e^{-(z + t/2) / \lambda_L}{sinh([z' - t/2] / \lambda_L) \over sinh(t
/ \lambda_L)}\right\}dz'\right ]dz \]
\begin{equation}
= -(H_\parallel - n_\parallel \Phi_o) \left\{
1 - {2\lambda_L \over t} tanh({t \over 2\lambda_L}) \right\} ,
\label{eq2}
\end{equation}
where n$_\parallel$ is the component of vortex density parallel to the 
surface.  The magnetic field generated parallel to the surface is
$n_\parallel \Phi_{eff}$ where,
\begin{equation}
\Phi_{eff} = \Phi_o \left\{ 1 - {2\lambda_L \over t} tanh({t \over 2\lambda_L}) \right\},
\label{eq3}
\end{equation}
is the average effective flux quantum of a vortex that is screened by image 
fields due to the surfaces.  For this sample, $\Phi_{eff} = 0.37 \Phi_o$.
Fig.\ \ref{fig7} (a) shows the magnetization obtained from the SPNR results of 
Fig.\ \ref{fig5} using Eq.\ \ref{eq2}.

In the SQUID magnetometer the magnetization is measured along the 
direction of the applied field so when the sample surface is slightly tilted from
the applied field by an angle $\theta$, a projection of the components 
of magnetization parallel and perpendicular to the surface will be measured:
\begin{equation}
M_\parallel cos\theta + M_\perp sin\theta.
\label{eq4}
\end{equation}

This is demonstrated quantitatively for our data in the low-field 
Meissner regime, below H$_{c1}$.  In this regime, the 
magnetization of a superconducting film at an angle with the applied 
field will be:
\begin{equation}
4 \pi M_o =  -H cos^2 \theta\left\{ 1- {2\lambda_L \over t} tanh({t \over
2\lambda_L}) \right\}- {H sin^2 \theta \over 1 - N}. 
\label{eq5}
\end{equation}
The inset to Fig.\ \ref{fig6} shows the slope of the 
low-field magnetic moment 
data of Fig.\ \ref{fig6} as a function of
tilt angle, where the solid curve is the result of a least-squares fit 
to Eq.\ \ref{eq5} 
with two free parameters, N and a scale factor.  
The latter permits an accurate determination of the magnetization, which 
also contains small instrumental effects \cite{mcelfresh}. 
As can be seen, the curve represents the data quite well and we obtain N 
= 0.998 $\pm$ 0.0015, which is slightly larger than 
the value 0.994 estimated for our sample geometry \cite{craik}.

More generally, we can include the effect of vortices on the magnetization
measured with the SQUID by substituting Eqs.\ \ref{eq1} and \ref{eq2} 
into Eq.\ \ref{eq4}, which gives 
\begin{equation}
4 \pi M = 4 \pi M_o +  n_\parallel \Phi_{eff} cos \theta
+ {n_\perp \Phi_o sin \theta \over 1 - N}, 
\label{eq6}
\end{equation}
where M$_o$ is the Meissner term in the absence of vortices, 
given by Eq.\ \ref{eq5}.

The conclusion is that the magnetometer is highly sensitive to the 
component of magnetization perpendicular to the film.  Although 
this component diminishes with decreasing $\theta$,
the demagnetization 
factor causes the denominator of Eq.\ \ref{eq1} to be small.  Thus, the SQUID 
measurements can, even for very small angles of tilt, be dominated by the 
magnetization perpendicular to the film.
This effect is also much 
larger for the vortex contribution than in the Meissner regime, since,
the Meissner effect diminishes as $sin^2 \theta$ but the vortex contribution
diminishes more slowly, as $sin \theta$.
Therefore, we conclude that the irreversibility observed in the SQUID measurement is
due to M$_\perp$, since, the SPNR shows M$_\parallel$ to be reversible.

In order to quantitatively compare the SPNR and SQUID experiments, 
the SQUID magnetization was 
measured by following the same field history as the SPNR measurement,
as shown by the small data points in Fig.\ \ref{fig7} (b).
This was done for the sample tilt angle nominally set to zero.  
The open circles in Fig.\ \ref{fig7} (b) are the ascending field SPNR 
data replotted 
from Fig.\ \ref{fig7} (a).  As can be seen, the SPNR and SQUID measurements 
quantitatively give the same magnetization for the virgin ascending field,  
indicating that the magnetization and vortices are parallel to 
the film plane.  
Moreover, the quantitative agreement between the two data sets also indicates 
that sample tilt effects as well as any contribution from vortices 
perpendicular to the surface are insignificant for the virgin ascending field.

The SQUID and SPNR measurements differ only subsequent to the 
virgin magnetization,  
which suggests that a magnetic field due to vortices 
is generated out of the film plane as the applied field is reduced. 
Thus, for descending field, the data were analyzed using two 
components of magnetic field,
one parallel ($n_{\parallel} \Phi_{eff}$) and 
one perpendicular ($n_{\perp} \Phi_o$) to the film plane.
The parallel magnetic field is obtained directly from the SPNR measurement and 
it is desireable to obtain the perpendicular field from the SQUID data.  
However, as can be seen from 
Eq.\ \ref{eq6}, the perpendicular field contribution to the SQUID magnetization 
also depends on the sample tilt angle, 
which is not precisely known, so that an additional piece of information is required.  
It is found that good quantitative agreement between the SQUID and SPNR data can be
obtained by assuming that the sum of the parallel and perpendicular magnetic fields is 
equal to the maximum trapped field:  
$n_{\parallel max} \Phi_{eff} =  n_{\parallel} \Phi_{eff} + n_{\perp} \Phi_o$,
where $n_{\parallel max}$ is the maximum vortex density parallel to the film plane,
which occurs at the maximum applied field.  
The large solid data points in Fig.\ \ref{fig7} (b) show the SQUID magnetization
predicted from the SPNR data using this relationship and a tilt angle of 
0.12$^o$.
Note that this angle is essentially zero to within the experimental error 
of setting the angle; yet, its finite size qualitatively impacts 
the shape of the magnetization curve by determining the size of the hysteresis loop.
The solid square SPNR data point was measured at 600 Oe after cycling the
field to -3000 Oe.  In that case, the maximum density of vortices was assumed to be
the same as was measured by SPNR at 3000 Oe 
and this also gives good agreement with the SQUID data.
An alternative model where the total vectorial
magnetic field is conserved, corresponding to 
$(n_{\parallel max} 
\Phi_{eff})^2 =  (n_{\parallel} \Phi_{eff})^2 + (n_{\perp} \Phi_o)^2$, gave a
substantially poorer comparison between the SPNR and SQUID data, although the 
qualitative trend was similar.

The perpendicular magnetic field can also be obtained independent of such models 
if a specific tilt-angle is assumed.  In this case, we have chosen two angles, 0.5$^o$ 
and 0.12$^o$.  The former is twice the maximum that could have 
been set experimentally and the latter is the value obtained  
from the above model (which we take as the lower limit because
a smaller angle would unphysically give 
$n_{\perp max} \Phi_o > n_{\parallel max} \Phi_{eff}$).
The result of this analysis is given in  
Figure\ \ref{fig8} (a) which shows the individual magnetic field components.
Fig.\ \ref{fig8} (b) shows the resulting angle between the magnetic field and the
film plane, 
$\omega$ = tan$^{-1}(n_\perp \Phi_o / n_\parallel \Phi_{eff})$.

\section{Discussion and Conclusions}

\begin{figure}
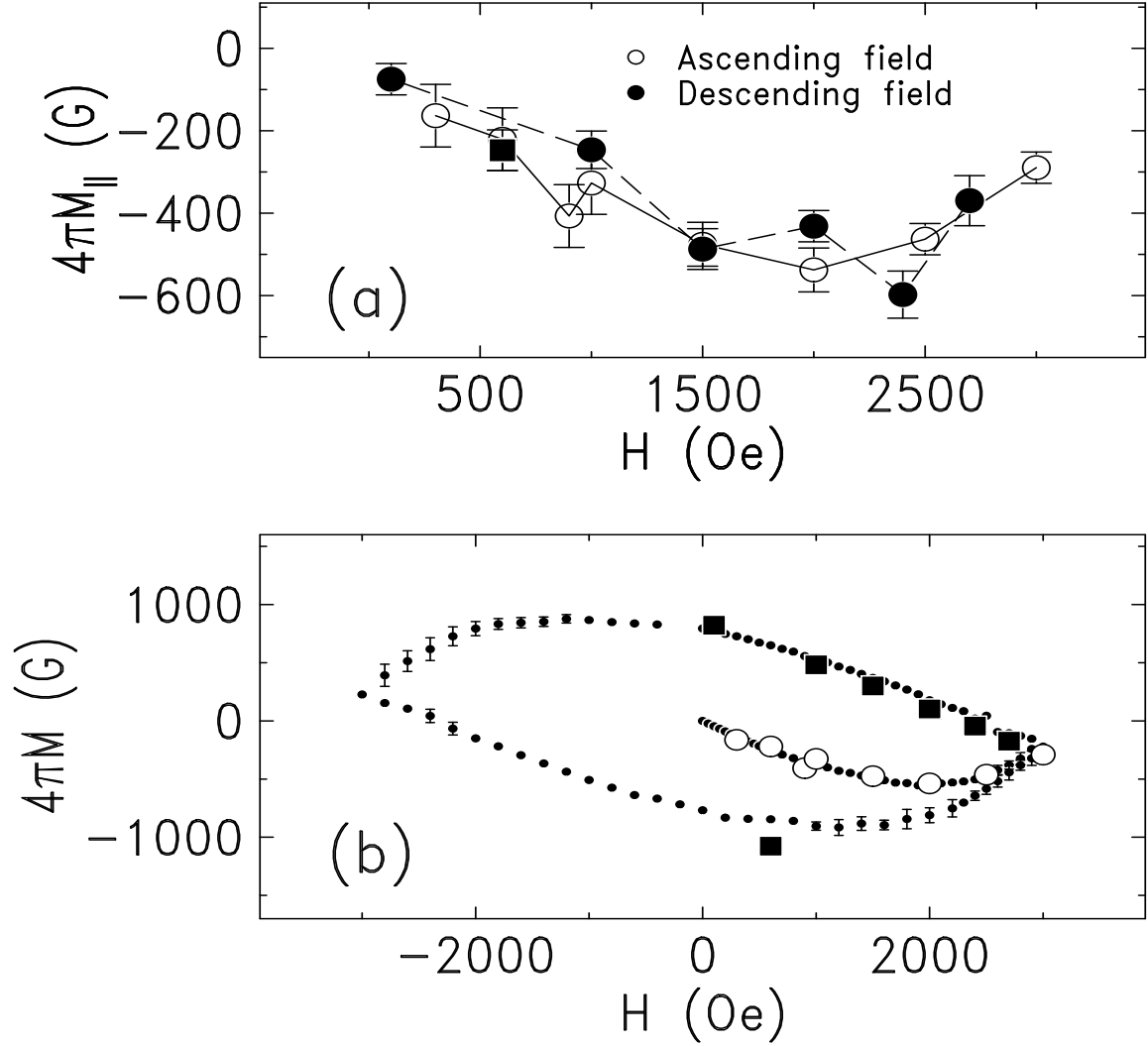

\caption{(a) The magnetization as a function of applied field 
is determined from the SPNR data in 
Fig.\ \ref{fig5} using Eq.\ \ref{eq2}. 
(b) Quantitative agreement between the SQUID and SPNR measurments is achieved
using the model described in the text.   
In (b) the small data points are the SQUID magnetization measured at 2 K
as a function
of applied field using the same field history as the SPNR experiment. 
The arrows indicate the order in which the data were collected.  
The open circles are the magnetization from the ascending field SPNR 
data replotted from (a), showing that the SPNR and SQUID measurements lead to
the same result:  for ascending field, the vortices are parallel to the film plane.  
The large solid circles are the descending field SPNR data converted 
to the conditions of the SQUID measurement assuming the 
model described in the text.  
The solid square comes from the SPNR data point measured after cycling the 
field to -3000 Oe.}
\label{fig7}
\end{figure}

\begin{figure}
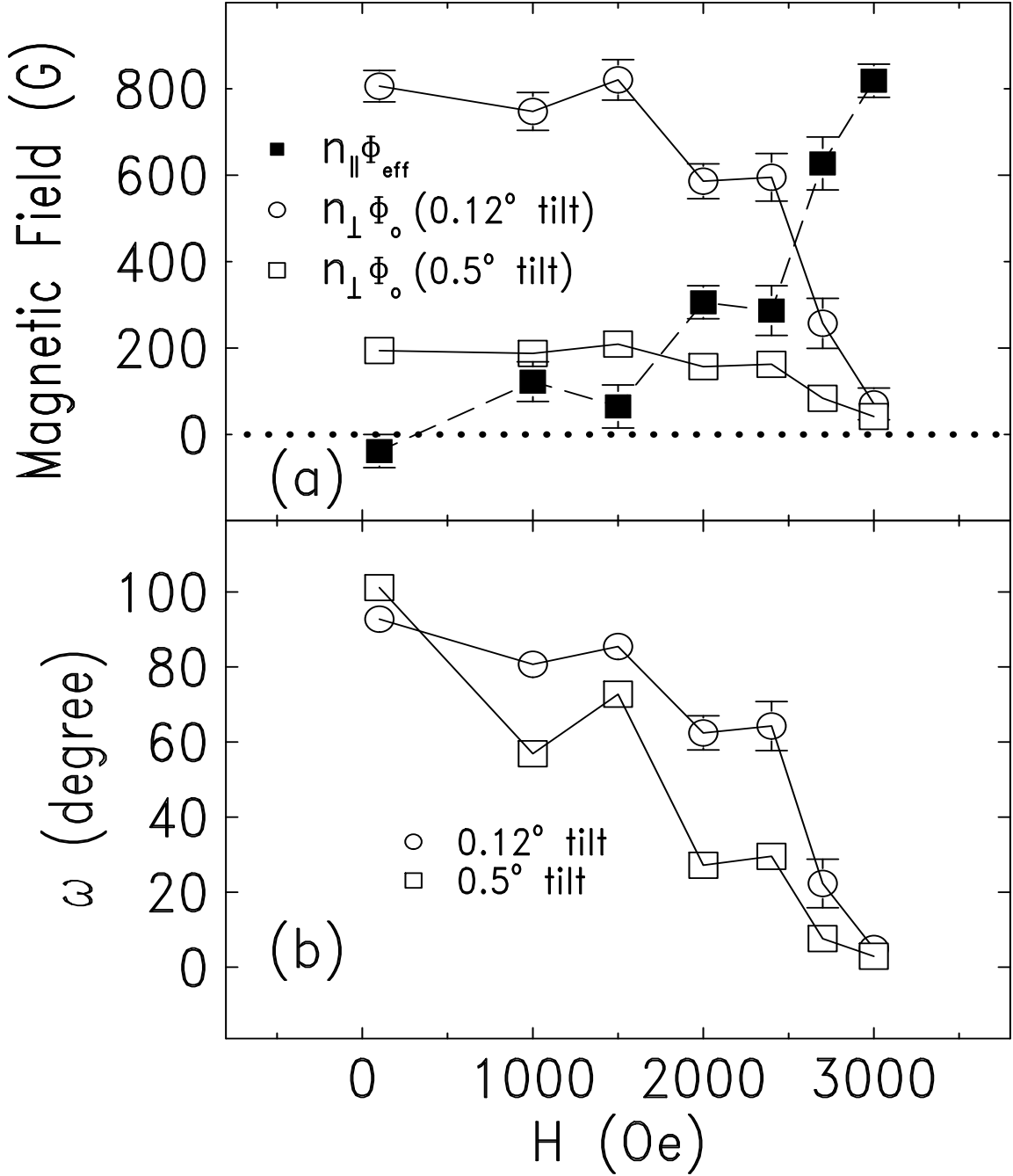

\caption{For descending applied field, the SQUID and SPNR measurements are used to 
determine the magnetic field in the sample. 
(a) The magnetic field parallel to the surface, given by the solid circles, is 
obtained directly from the SPNR measurement.
The perpendicular magnetic field, shown by the 
open circles (0.12$^o$) and open squares (0.5$^o$), is obtained by solving
Eq. (3.6) using the SQUID and SPNR data and by assuming the given sample tilt angles.
(b) shows the angle between the magnetic field and the surface for the two assumed
angles of sample tilt.}
\label{fig8}
\end{figure}

The good quantitative agreement between the SQUID and SPNR data using the above 
analysis supports the basic conclusion
that the initial magnetic field of the vortices is parallel to the film plane 
and it subsequently 
rotates perpendicular to the film plane as the applied field is removed.  
Although the analysis presented in Fig.\ \ref{fig7} (b) assumes that the 
trapped magnetic field is strictly conserved,
one cannot draw this conclusion because of the uncertainty 
in the sample tilt-angle.
Fig.\ \ref{fig8} demonstrates the range of perpendicular magnetic fields that 
can be obtained, which depend upon the assumed sample tilt. 
Assuming the maximum tilt leads to a lower limit 
on the maximum trapped perpendicular magnetic field:
$n_{\perp max} \Phi_o$ = 0.25 $n_{\parallel max} \Phi_{eff}$.
In this limiting scenario, 75\% of the magnetic field
leaves the sample when the applied field is reduced to zero.   
In any case, it is safe to conclude that a significant fraction of the 
magnetic field is trapped at remanance and that it is oriented perpendicular to 
the film plane.

Based on previous studies, these results are different from what one might expect.  
For example, our SPNR investigation \cite{han} of a YBCO thin-film 
showed a large hysteresis loop and remanence, indicating a strong preference for
the vortices to remain parallel to the film plane.  
From the smaller coherence length of YBCO one expects significantly 
stronger vortex pinning in YBCO than Nb.  However, the rotation of the 
trapped magnetic field
in Nb rather than the removal of vortices is not expected.  
Other studies of anisotropic
superconductors \cite{hess} have found that the vortex density perpendicular to the 
surface is essentially proportional to the projection of the applied field.
However, those studies were performed on thick crystals and the vortices were 
examined upon field-cooling.
Given the isotropy of Nb,
we expect that the observed magnetic field rotation is due to the thin-film
nature of the sample (thickness/$\lambda_L \sim$2.8). 

We can speculate on two surface-related mechanisms that give 
a lower equilibrium 
energy for vortices oriented perpendicular to the film plane.  
First, the vortex density in the perpendicular direction is 
no more than 37$\%$ of the maximum parallel vortex density 
(because $n_{\perp max} \Phi_o  \le n_{\parallel max} \Phi_{eff}$)
and this leads to a lower vortex-vortex repulsion for vortices
reoriented perpendicular to the film plane.  
A second mechanism relates to the weaker vortex binding energy 
(one-body interaction) for the parallel geometry where
there are two energy contributions
arising from the surface \cite{degennes}: a repulsive term due to the 
vortex interaction with the surface-screening field and an attractive term due to
the interaction with the image vortex.  The repulsive term is proportional to H and 
will dominate at higher field, thereby leading to smaller 
binding energy for vortices parallel to the surface.  
Whether these mechanisms are relevant will depend on kinetics as well as crystalline
anisotropy.  It could be that the latter plays a role in the 
YBCO thin-film \cite{han}.

Also of interest are the pinning, rotation, and nucleation behavior 
of vortices in this geometry. 
One might expect, to a first approximation, that the mechanism for vortex rotation
is similar to vortex removal:  the vortex line must move towards a surface.  
However, the vortices appear to rotate rather than exit the sample, suggesting
that opposite ends of a vortex line penetrates the opposite planar surfaces of
the film.  
Once rotated, the vortex pinning is much stronger in the perpendicular geometry
\cite{zeldov}.  
Vortex rotation alone, however, does not explain these results.  
Given the magnitude of the perpendicular magnetic field found in 
Fig.\ \ref{fig8}, many additional
vortices must be generated when the field is reduced because the planar area
of the film is nearly $10^5$ times larger than the area of the film's edge.
It might be that, with decreasing applied field, the segment of a 
vortex that is parallel to the surface decreases in length 
(effectively giving a rotation, since, the "threading" segment perpendicular
to the surface is constant in length) 
while, simultaneously, new vortices are nucleated in order 
to increase the number of vortex threads that penetrate the planar surface.
Clearly, further theoretical and experimental work is necessary in order
to gain a better understanding of the 
vortex behavior in this geometry.

Support is gratefully acknowledged (PFM, SWH, JF) from the Midwest
Superconductivity Consortium (MISCON) under DoE grant DE-FG02-90ER45427,
the NSF DMR 96-23827, and (LHG, IVR) from DoE Grant DE FG02 96 ER45439.
The SPNR experiments were performed at the Missouri University Research Reactor (MURR).

\begin{references}
\bibitem[**]{paul} Author to whom correspondence should be addressed.
\end{references}
\end{document}